\documentclass[aps,prb,reprint,groupedaddress]{revtex4-1}
\usepackage[linktocpage,colorlinks,breaklinks,linkcolor=blue,citecolor=blue,urlcolor=blue]{hyperref}
\usepackage{graphicx}
\usepackage{amsmath}
\usepackage{color}
\usepackage[normalem]{ulem}

\definecolor{purple}{rgb}{0.7,0.0,0.7}
\definecolor{orange}{rgb}{1,0.65,0.0}

\begin{document}


\title{Enhancement models of momentum densities of annihilating electron-positron pairs: the many-body picture of natural geminals}


\author{Ilja Makkonen}
\email[]{ilja.makkonen@aalto.fi}
\affiliation{COMP Centre of Excellence and Helsinki Institute of Physics, Department of Applied Physics, Aalto University School of Science, P.O.\ Box 11100, FI-00076 AALTO, Finland}
\author{Mikko M. Ervasti}
\affiliation{COMP Centre of Excellence and Helsinki Institute of Physics, Department of Applied Physics, Aalto University School of Science, P.O.\ Box 11100, FI-00076 AALTO, Finland}
\author{Topi Siro}
\affiliation{COMP Centre of Excellence and Helsinki Institute of Physics, Department of Applied Physics, Aalto University School of Science, P.O.\ Box 11100, FI-00076 AALTO, Finland}
\author{Ari Harju}
\affiliation{COMP Centre of Excellence and Helsinki Institute of Physics, Department of Applied Physics, Aalto University School of Science, P.O.\ Box 11100, FI-00076 AALTO, Finland}


\date{\today}

\begin{abstract}
The correlated motion of a positron surrounded by electrons is a fundamental many-body problem. We approach this by modeling the momentum density of annihilating electron-positron pairs using the framework of reduced density matrices, natural orbitals and natural geminals (electron-positron pair wave functions) of the quantum theory of many-particle systems. We find that an expression based on the natural geminals provides an exact, unique and compact expression for the momentum density. The natural geminals can be used to define and determine enhancement factors for enhancement models going beyond the independent-particle model for a better understanding of results of positron annihilation experiments. 
\end{abstract}


\maketitle


When a positron annihilates with an electron, the emitted $\gamma$ photons
provide valuable information on the surroundings in which they annihilate.
This is used in the positron emission tomography for living subjects,\cite{Gambhir2002}
and in annihilation spectroscopy for detection,
quantification, and chemical and structural characterization of
open-volume defects in materials~\cite{TuomistoRMP} and for studying
Fermi surface
of metals and alloys.~\cite{West1993,*Mijnarends1993} In positron annihilation experiments, positrons entering a solid thermalize very rapidly and trap effectively at open-volume defects, in which their measured lifetimes are increased.
Furthermore, the momentum density of annihilating electron-positron
pairs can reveal impurity or dopant atoms around the trap or give
information on the structure of the Fermi surface and correlation
effects reflected in its shape.

Theoretical modeling has an important
role in understanding the indirect information contained in the
experimental results.~\cite{TuomistoRMP} The measurable quantities can be modeled using
electronic structure techniques, but typically instead of using direct
many-body modeling of positrons in a solid (see, for example,
Ref.~\onlinecite{Sormann1996}) one has to resort to using
a mean-field approach for the electrons and positrons.
Much of the modeling in the defect identification field is based on the
two-component electron-positron density functional theory~\cite{Boronski1986} (DFT), but in this formalism the fundamental quantities
are the real-space one-body electron and positron densities,
and the route to obtain two-body momentum-space observables is not a practical
one.~\cite{Bauer1983,*Barbiellini1999} 
In practice, when modeling the momentum densities one has to resort to band-structure calculations and choose one of the various modifications of the
independent-particle model (IPM),
\begin{equation}\label{IPM}
 \rho(\mathbf{p})=\sum_{\text{occ.}\ j}\left|\int d\mathbf{r}\,
 e^{-i\mathbf{p}\cdot\mathbf{r}}\phi^{p}(\mathbf{r})\phi^{e}_{j}(\mathbf{r})\right|^{2},
\end{equation}
appropriate for non-interacting systems (see, for example
Refs.~\onlinecite{Mijnarends1979,Daniuk1987,*Jarlborg1987,AlataloPRB1996,Laverock2010}).
What the IPM does not describe is the effect of short-range screening
of the positron by electrons, a many-body effect increasing the
annihilation rate and affecting the measured $\rho(\mathbf{p})$. In some models,\cite{Daniuk1987,*Jarlborg1987} this phenomenon is incorporated by scaling the orbital product $\phi^{p}(\mathbf{r})\phi^{e}_{j}(\mathbf{r})$ by parametrized enhancement factors as $\sqrt{\gamma_j (\mathbf{r})} \phi^{p}(\mathbf{r})\phi^{e}_{j}(\mathbf{r})$, to describe the correlated electron-positron pair wave function more realistically.

How exactly should the IPM be modified in order to take correlations into account, preferably both electron-positron and electron-electron
ones? This way one could calculate accurate momentum densities with the two-component density-functional theory without the need of expensive many-body modeling. To answer this, we utilize concepts of reduced density matrices (RDM), natural
orbitals (NO's), and natural geminals.
\cite{Lowdin1955,Coleman1963} 
The many-body reduced density-matrix formalism combined with many-body
modeling can be used to derive and analyze expressions, which resemble
the IPM of Eq.~(\ref{IPM}) but are exact even for correlated
systems. The reduced density matrices and NO's have been discussed earlier in the
context of electron momentum densities and X-ray Compton scattering,\cite{Barbiellini2000,*Barbiellini2001} where they are
intriguing since for the NO's the exact expression is simple and similar to the analogous IPM for the electron momentum density. Below we demonstrate that
the two-body objects corresponding to the NO's in case of electron-positron
annihilation are the natural geminals.\cite{Coleman1963} They provide an exact and unique connection from the many-body picture to the two-particle one of the annihilating pair
and lead to a compact formula for
$\rho(\mathbf{p})$, in which we can further connect the natural geminals to products of single-particle orbitals (NO's or the orbitals within DFT and the Kohn-Sham method) using enhancement factors describing the correlated motion. The resulting expression still has a solid basis in the many-body formalism.
We perform accurate many-body modeling of finite inhomogeneous electron-positron systems
using the exact diagonalization (ED) technique to get accurate
reference momentum densities and to benchmark the suggestion to go
beyond the IPM.

We focus on the $2\gamma$ annihilation in which the annihilating
electron-positron pair is in spin singlet state. The momentum density
of the annihilating pairs reads
\begin{equation}\label{momdens}
 \rho(\mathbf{p})\propto\int d\mathbf{r}d\mathbf{r}'\,e^{-i\mathbf{p}\cdot(\mathbf{r}-\mathbf{r}')}\Gamma^{ep}(\mathbf{r},\mathbf{r};\mathbf{r}',\mathbf{r}') ,
\end{equation}
where $\Gamma^{ep}(\mathbf{r},\mathbf{r};\mathbf{r}',\mathbf{r}')$ is the electron-positron two-body reduced density matrix (2-RDM), defined as
\begin{align}\label{2RDM}
 &\Gamma^{ep}(\mathbf{r}_{p},\mathbf{r}_{e};\mathbf{r}_{p}',\mathbf{r}_{e}')
 =N_{e}N_{p}\int d\mathbf{r}_{3}\ldots d\mathbf{r}_{N}\nonumber\\
 &\times\Psi(\mathbf{r}_{p},\mathbf{r}_{e},\ldots,\mathbf{r}_{N})\Psi^{*}(\mathbf{r}_{p}',\mathbf{r}_{e}',\ldots,\mathbf{r}_{N}).
\end{align}
Here, $\Psi(\mathbf{r}_{1},\mathbf{r}_{2},\ldots,\mathbf{r}_{N})$ is
the many-body wave function of the whole system, $N_{e}$ and $N_{p}$ are the particle numbers, $\mathbf{r}_{i}$ the position of particle $i$, and the subscripts $e$ and $p$ denote electrons and positrons with fixed, opposite spin components. 
We assume the usual experimental case of only one positron in the system at a time.

In the many-body formalism, especially within second quantization, the 2-RDM is typically expanded using single particle orbital bases, say $\{\phi^{e}_{i}\}$ and $\{\phi^{p}_{i}\}$, such that
\begin{equation}\label{orbitalexp}
\Gamma^{ep}(\mathbf{r}_{p},\mathbf{r}_{e};\mathbf{r}_{p}',\mathbf{r}_{e}') = \sum_{ijkl} \rho_{ijkl}^{ep} {\phi_{i}^{p}}^*(\mathbf{r_p'}) {\phi_{j}^{e}}^*(\mathbf{r_e'}) {\phi_{k}^{p}}(\mathbf{r_p}) {\phi_{l}^{e}}(\mathbf{r_e}),
\end{equation}
where the factors are $\rho_{ijkl}^{ep} = \langle a^{\dagger}_{ip}a^{\dagger}_{je}a_{le}a_{kp} \rangle$. The IPM, which is the exact result for a noninteracting system, results from this expression. However, regardless of the orbital bases used, it will turn out below that for an interacting system there are too many significant coefficients $\rho_{ijkl}^{ep}$ to construct a parametrization or a meaningful model that takes correlations into account. Therefore, instead of trying to express the 2-RDM using single-particle functions, one can expand by electron-positron pair wave functions $\{w_i\}$, also called geminals,
\begin{equation}\label{geminalexp}
\Gamma^{ep}(\mathbf{r}_{p},\mathbf{r}_{e};\mathbf{r}_{p}',\mathbf{r}_{e}')=\sum_{ij} b_{ij} \omega_{i}(\mathbf{r}_{p},\mathbf{r}_{e}) \omega_{j}^{*}(\mathbf{r}_{p}',\mathbf{r}_{e}').
\end{equation}
Particularly useful are the natural geminals $\{\alpha_j\}$, which are the unique orthonormal eigenfunctions of the 2-RDM \cite{Coleman1963}
\begin{equation}\label{eigeq}
 \int d\mathbf{r}_{p}'d\mathbf{r}_{e}'\,\Gamma^{ep}(\mathbf{r}_{p},\mathbf{r}_{e};\mathbf{r}_{p}',\mathbf{r}_{e}')\alpha_{j}(\mathbf{r}_{p}',\mathbf{r}_{e}')=g_{j}\alpha_{j}(\mathbf{r}_{p},\mathbf{r}_{e}).
\end{equation}
The eigenvalues $g_{j}$ are non-negative occupations and satisfy $\sum_j g_j = N_p N_e$. Furthermore, the natural geminals provide a diagonal expansion in Eq.~(\ref{geminalexp}).

Expanding using the natural geminals, the momentum density of the annihilating pairs, Eq.~(\ref{momdens}), becomes
\begin{equation}\label{geminalmd}
 \rho (\mathbf{p})\propto\sum_{j}g_{j}\left |\int d\mathbf{r}\,e^{-i\mathbf{p}\cdot\mathbf{r}}\alpha_{j}(\mathbf{r},\mathbf{r})\right |^{2} ,
\end{equation}
where, unlike in the corresponding orbital expansion based expression [see Eq.~(\ref{helikopteri}) below], each term is positive at all momenta $\mathbf{p}$.
Similar diagonal expressions have appeared in the literature (see, for example, Ref.~\onlinecite{Sormann2006}), where the geminals have been denoted as ``electron-positron pair wave functions''. In the present work we are for the first time able to discuss their nature in a general interacting many-body system and identify them unambiguously as natural geminals. One can possibly relate the natural geminals to the Lehmann representation geminals of the two-particle Green's function similarly as the natural orbitals are connected to the generalized overlap amplitudes.\cite{Goscinski1970}

In addition to the momentum density of annihilating pairs, another important experimental parameter is the positron annihilation rate.
With all four coordinates the same, the above 2-RDM is the contact density, namely the density of electrons at a positron at point $\mathbf{r}$, the quantity determining the local positron annihilation rate,
$\gamma(\mathbf{r})n_{e}(\mathbf{r})n_{p}(\mathbf{r})=\Gamma^{ep}(\mathbf{r},\mathbf{r};\mathbf{r},\mathbf{r})$.
Here $\gamma(\mathbf{r})$ is the so-called enhancement factor of the total density (the zero-distance value of the electron-positron pair correlation function at $\mathbf{r}$) appearing in the positron literature. Its purpose is to incorporate many-body effects, namely to take the short-range screening of the positron by the electrons into account when evaluating the positron annihilation rate $\lambda$ or the mean lifetime $\tau$ using the DFT's uncorrelated one-body electron and positron densities $n_{e}(\mathbf{r})$ and $n_{p}(\mathbf{r})$,
\begin{equation}
 \lambda=\frac{1}{\tau}\propto\int d\mathbf{r}\,\gamma(\mathbf{r})n_{e}(\mathbf{r})n_{p}(\mathbf{r}).
\end{equation}
Typically $\gamma (\mathbf{r})$ is approximated within the local-density approximation (LDA) and the parametrizations are based on many-body calculations made for homogeneous electron-positron systems.\cite{Boronski1986,Puska1995,Barbiellini1995,Drummond2011} How to define the enhancement factor $\gamma (\mathbf{r})$ for the total particle densities is straightforward. The next question would be how to properly define similar enhancement factors for correcting momentum densities and products of uncorrelated single-particle orbitals.

The natural geminal expansion Eq.~(\ref{geminalmd}) differs from the IPM of Eq.~(\ref{IPM}) by the eigenvalues $g_j$ and that the product of the single-particle orbitals is replaced by a geminal involving correlations. 
In the limit of vanishing electron-positron interaction, the natural geminals are simply products of the positron and electron natural orbitals $\varphi^{p}_0(\mathbf{r})$ and $\varphi^{e}_j(\mathbf{r})$ (normalized eigenfunctions of the respective 1-RDM's), $\alpha_{j}^{0}(\mathbf{r},\mathbf{r})=\varphi^{p}_0(\mathbf{r})\varphi^{e}_j(\mathbf{r})$, and the eigenvalues $g_j$ are the products of the electron and positron natural orbital occupations (eigenvalues of the 1-RDM). Therefore, and to meet and interpret the state and position-dependent enhancement factors of enhancement models,\cite{Daniuk1987,*Jarlborg1987} we define $\sqrt{\gamma_{j}(\mathbf{r})}$ using
\begin{equation}\label{enhancement}
\alpha_j(\mathbf{r},\mathbf{r}) = \sqrt{\gamma_j(\mathbf{r})} \varphi_{0}^{p}(\mathbf{r}) \varphi_j^{e}(\mathbf{r}) .
\end{equation}
Hence, the many-body interpretation of the state and position dependent enhancement factor is to relate natural orbitals to natural geminals. This can be made explicit by expanding the geminals by natural orbitals $\alpha_j(\mathbf{r}_p,\mathbf{r}_e) = \sum_{mn} c_{mn}^{(j)} \varphi_m^{p}(\mathbf{r}_p) \varphi_n^{e}(\mathbf{r}_e)$, where the dominant term should be $\varphi^{p}_0(\mathbf{r}_p) \varphi_j^{e}(\mathbf{r}_e)$ for the enhancement factor to be reasonable. This seems to be true for natural geminals of high occupation $g_j$, but for low $g_j$ the Schmidt decomposition shows that the natural geminals are entangled with orbitals that are more deformed from natural orbitals.
Furthermore, the nodes of the natural geminals and natural orbitals do not match in general, in which case the defined enhancement factor can exhibit singularities. We have not encountered such problems in our model systems.

In our model system, we confine eight electrons and one positron in a
three-dimensional harmonic trap,
$V_{\text{ext}}(\mathbf{r})=\omega^{2}r^{2}/2$ (we use the Hartree atomic units).
\cite{[{
For a more realistic model, see }]{Saniz2002}}
The exact
diagonalization (ED) method is used to solve the ground
state, with the many-body basis truncated by allowing
only Slater determinants of the lowest 8 non-interacting
shells (120 orbitals) with non-interacting total energy less than a chosen
cutoff ($E\le 33 \omega$). The 1-RDM and 2-RDM are straightforward to extract from the total wave function, and natural orbitals and geminals are solved by diagonalizing these.
Our model systems have the two lowest electron shells (4 orbitals) occupied within IPM.
The magnitudes and degeneracies of the largest $g_{j}$ and the symmetries of the corresponding $\alpha_{j}(\mathbf{r},\mathbf{r})$ reflect the very same shell structure seen in either the NO's and their occupations 
or in DFT's LDA orbitals and eigenvalues. To exemplify, for a $\omega=0.4$ system with an electron density parameter $r_{s}\ge1.6$~a.u.\ we find a singly degenerate $g_{0}=0.943$ and a triply degenerate $g_{j}=0.917$ ($j=1,2,3$), while regardless of the $\omega$ value the corresponding $\alpha_{j}(\mathbf{r},\mathbf{r})$ have $s$ and $p$ type symmetries arising from products of $s$ and $p$ type electron orbitals with an $s$ type positron orbital. 
Moreover, according to our data the enhancement factors for our spherically symmetric models are shell-dependent radial functions. This suggests that one could be able to parametrize them as local or semi-local (shell) density functionals.

\begin{figure*}[t]
\begin{minipage}[t]{0.38\linewidth}
\includegraphics[width=\textwidth]{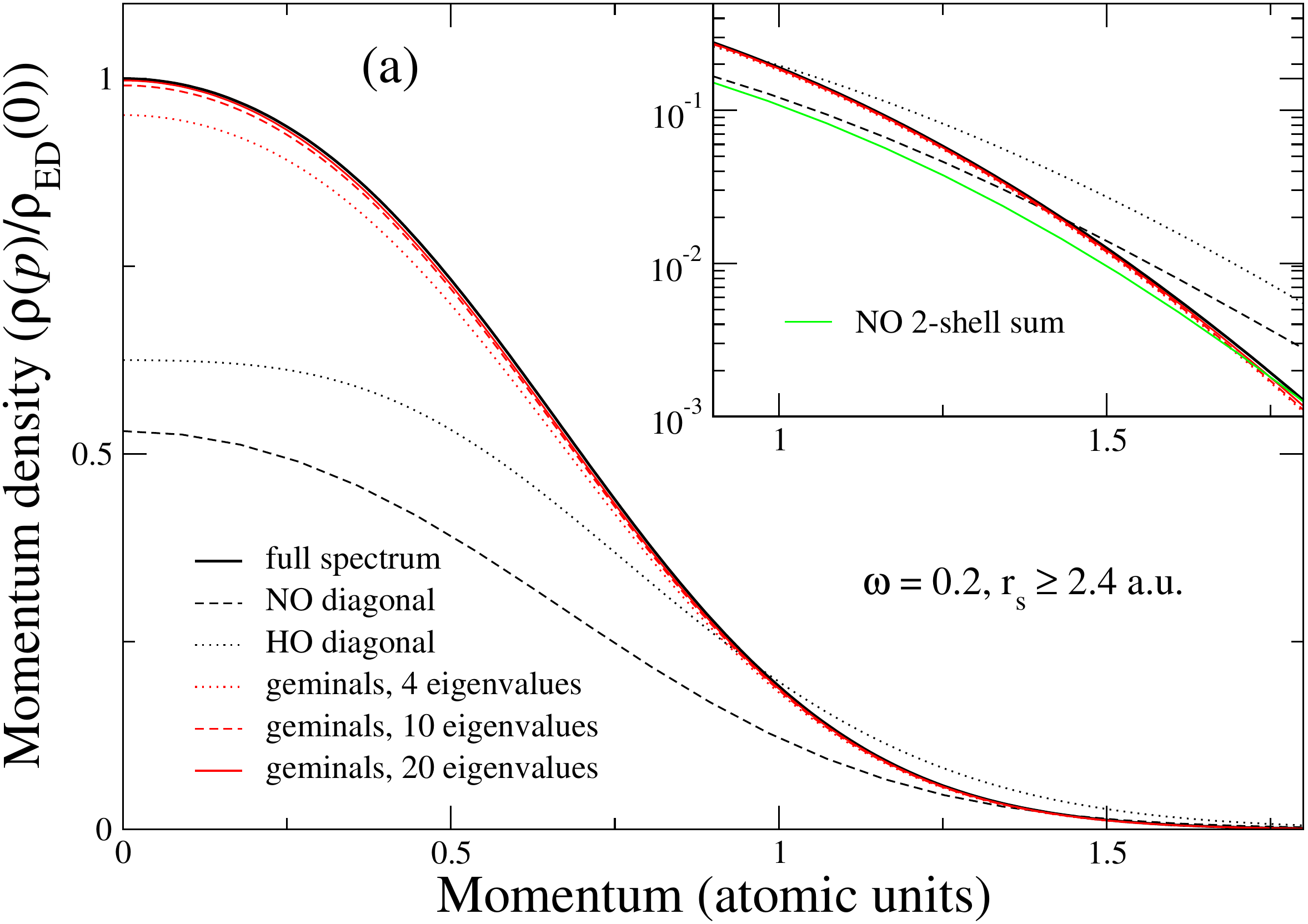}%
\end{minipage}
\hfill
\begin{minipage}[t]{0.205\linewidth}
\includegraphics[width=\textwidth]{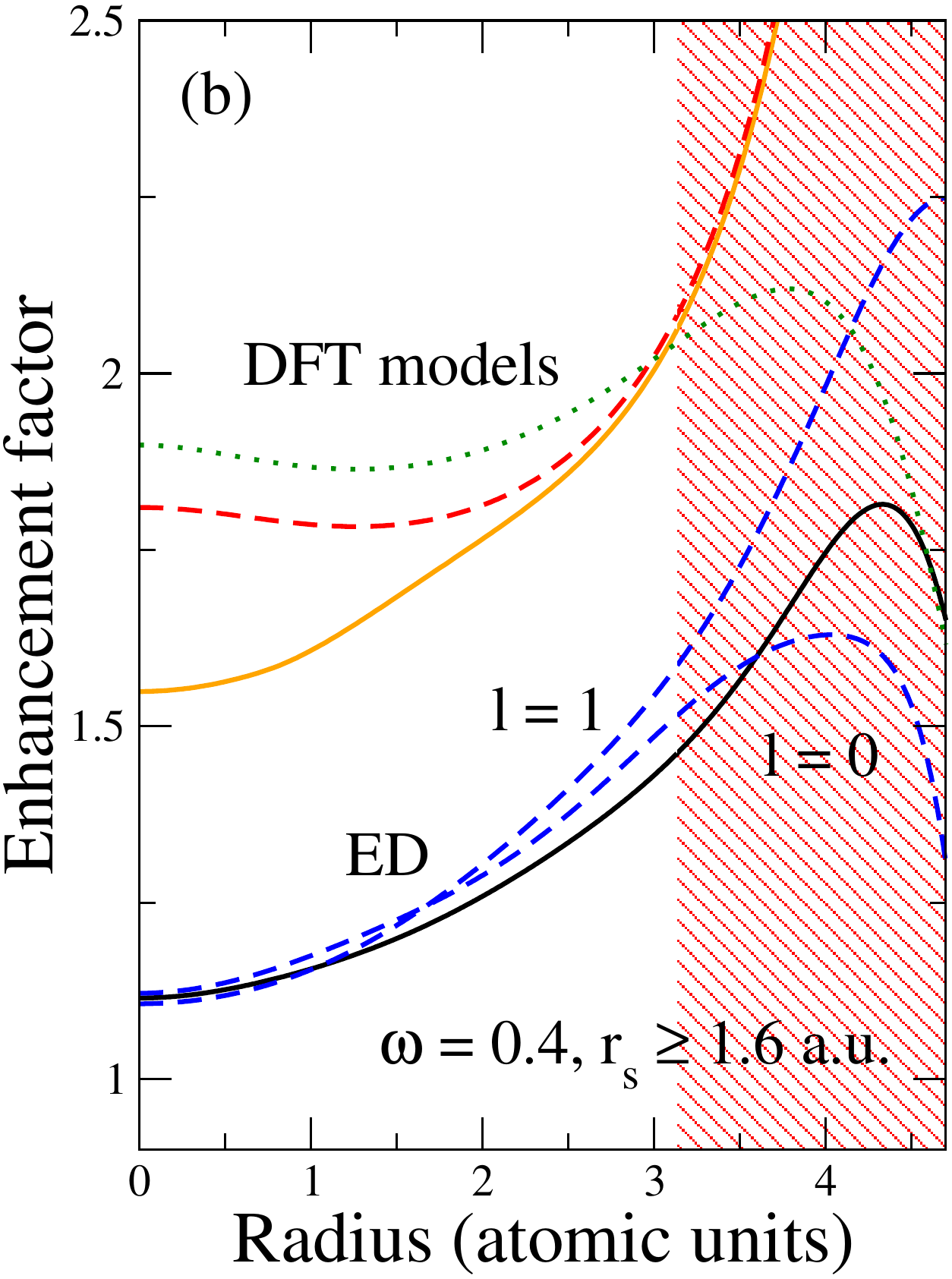}%
\end{minipage}
\hfill
\begin{minipage}[t]{0.38\linewidth}
\includegraphics[width=\textwidth]{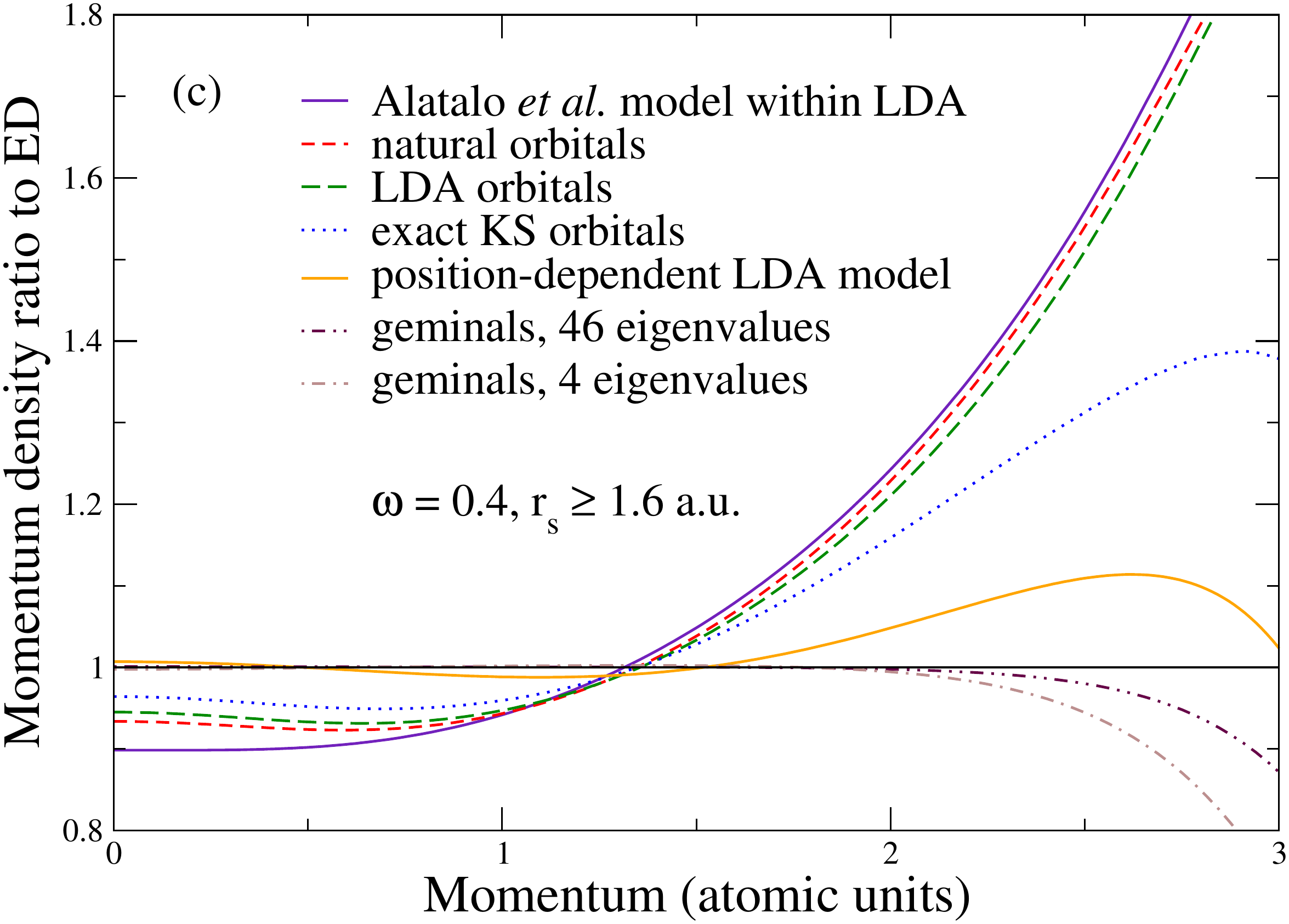}%
\end{minipage}
\caption{(Color online) (a) The total momentum density, unnormalized diagonal sums of Eq.~(\ref{helikopteri}) for the NO and HO orbitals, and the convergence analysis of Eq.~(\ref{geminalmd}) for the $\omega=0.2$ system. The inset shows the high momentum regime and the behavior of the 2-shell NO sum. (b) Enhancement factors for the $\omega=0.4$ system. The solid (black) line shows the factor $\sqrt{\gamma (\mathbf{r})}$ evaluated with ED. The (blue) dash lines show the $\sqrt{\gamma_{j} (\mathbf{r})}$'s for the $s$ and $p$ shells. The solid (orange) line is the $\sqrt{\gamma (\mathbf{r})}$ according to a two-component LDA functional,\cite{Puska1995} the (red) dash line its zero-positron-density limit,
\cite{Boronski1986} and the dotted (green) one the result of the GGA (Ref.~\onlinecite{Barbiellini1995}), which is also formulated in this limit. The ED results are not converged near the border of the figure because of the finite extent of the basis functions.
(c) 2-shell diagonal sums of Eq.~(\ref{helikopteri}) for various bases as well a convergence analysis of Eq.~(\ref{geminalmd}) for the $\omega=0.4$ system plotted as a ratio to the accurate ED spectrum. All spectra have been normalized prior to taking the ratio.
Also the position-dependent LDA model employing the LDA $\gamma (\mathbf{r})$ (Ref.~\onlinecite{Daniuk1987,*Jarlborg1987}) has been applied as well as the state-dependent enhancement factor by \citet{AlataloPRB1996}
\label{compositefigure}}
\end{figure*}

 We start our analysis of momentum densities calculated for our model systems by considering the convergence of the natural geminal expansion expression. Figure~\ref{compositefigure}(a) shows the full momentum density and the convergence of  Eq.~(\ref{geminalmd}) as a function of the number of terms included for an $\omega=0.2$ system with $r_{s}\ge 2.4\ a_{0}$.  We observe a rapid convergence of both the shape and the mass of the spectrum. The 4 eigenvalues corresponding to the pairs of orbitals occupied in the IPM are not quite enough to reproduce the mass but the shape is already good. The natural geminal expansion expression appears to be much more compact than the corresponding single-particle orbital expression based on Eq.~(\ref{orbitalexp}), 
\begin{align}\label{helikopteri}
 \rho(\mathbf{p})\propto
 \sum_{ijkl}&\rho_{ijkl}^{ep}\left[\int d\mathbf{r}\,e^{-i\mathbf{p}\cdot\mathbf{r}}\phi^{p}_{k}(\mathbf{r})\phi^{e}_{l}(\mathbf{r})\right] \nonumber\\
 &\times \left[\int d\mathbf{r}'\,e^{-i\mathbf{p}\cdot\mathbf{r}'}\phi^{p}_{i}(\mathbf{r}')\phi^{e}_{j}(\mathbf{r}')\right]^{*} , 
\end{align}
namely the important off-diagonal terms of the 4-index sum of Eq.~(\ref{helikopteri}) are reproduced by the natural geminal terms in the 1-index sum of Eq.~(\ref{geminalmd}). To be useful for real calculations, Eq.~(\ref{helikopteri}) should have rather diagonal $\rho_{ijkl}^{ep}$ in a convenient basis and converge fast as a function of number of terms included. This appears not to be the case as will be demonstrated below.

We test Eq.~(\ref{helikopteri}) by restricting to diagonal sums consisting of terms corresponding to $\rho^{ep}_{ijij}$ and studying how they look for different choices of the bases.
We choose (\emph{i}) NO's \cite{Lowdin1955} calculated with ED, orbitals (\emph{ii}) calculated using
fully self-consistently the two-component electron-positron
density-functional theory \cite{Boronski1986} within the LDA for the
electron-electron exchange and correlation energy \cite{Perdew1981} and
electron-positron correlation energy,\cite{Puska1995} (\emph{iii}) 
obtained using accurate ED densities and an inversion algorithm providing the corresponding ``exact'' local potentials and orbitals,\cite{vanLeeuwen1994,Makkonen2012}
and (\emph{iv}) those of a
noninteracting harmonic oscillator (HO). 
%

Figure~\ref{compositefigure}(a) displays also sums of the diagonal terms of Eq.~(\ref{helikopteri}) with the factors $\rho^{ep}_{ijij}$ from ED for two exemplary basis sets, the NO and the HO
orbitals. The discrepancy between the truncated sums and the full results
in Fig.~\ref{compositefigure}(a) clearly shows that the off-diagonal
terms do, regardless of the basis used, have a significant
contribution to the mass of the full spectrum (although not always to
their shape) 
especially in strongly interacting
systems (small $\omega$) such as this one. Since the interactions are reflected better in the NO's, the shape of the spectrum is
reproduced better than with HO orbitals. The same applies also to the
KS orbital bases (not shown). The mass of the full spectrum is larger
than that of the diagonal sum, which corresponds to an increasing
positron annihilation rate with increasing number of off-diagonal
terms. The inset of Fig.~\ref{compositefigure}(a) inspects the
high-momentum region of the spectra in case of the NO basis. The
diagonal sum has a larger intensity at high momenta than the full one,
especially in case of this more strongly interacting system. This kind of a convergence is possible for Eq.~(\ref{helikopteri}) since there is no guarantee that the off-diagonal elements of $\rho^{ep}_{ijkl}$ or the corresponding terms would be positive.
It turns
out that this overestimation can be cured by including in the sum only
the terms corresponding to states, which are occupied in the IPM, namely the lowest positron state and two occupied electron shells [see the inset of
Fig.~\ref{compositefigure}(a)]. We have analyzed the convergence as a function of the  number of off-diagonal terms using the HO basis, and the convergence seems rather slow, too slow to provide a simple model for improving the 2-shell diagonal sums. The spectra in
Fig.~\ref{compositefigure}(a) are unnormalized. However, when they are
normalized to the same number of ``counts'' as when comparing theory
against experiments, the discrepancy appears smaller. 

We have now established the rapid convergence of the natural geminal expansion in comparison with the poor performance of the orbital-based formula. The next step is then to demonstrate in practice how the many-body problem can be connected in another way with the single-particle picture using the natural geminals and enhancement factors as in Eq.~(\ref{enhancement}).
Figure~\ref{compositefigure}(b) shows a comparison of different enhancement factors for the $\omega=0.4$ system with $r_{s}\ge 1.6\ a_{0}$. We use the exact limit, the NO's as our reference orbitals. The link to the practical DFT calculations here is that they differ very little from the KS orbitals. We evaluate from our ED data the square root of the enhancement factor of the total density $\sqrt{\gamma (\mathbf{r})}$, and the state-dependent enhancement factors of Eq.~(\ref{enhancement}). The agreement between these quantities is rather good, which provides some support for the use of 
the accurate total density's $\gamma (\mathbf{r})$ in models with a position-dependent enhancement,\cite{Daniuk1987,*Jarlborg1987} although some state-dependence can clearly be seen. Figure~\ref{compositefigure}(b) also compares the above quantities to different LDA enhancement factors evaluated from LDA densities. The two-component LDA parametrization of Ref.~\onlinecite{Puska1995} reproduces the correct shape but the magnitude is too large, whereas the zero-positron-density limit appropriate for delocalized positrons has within the LDA (Ref.~\onlinecite{Boronski1986}) also a wrong shape. This applies also to the generalized-gradient approximation\cite{Barbiellini1995} (GGA). The structure in these two $\sqrt{\gamma (\mathbf{r})}$'s arises from having a non-monotonous electron density profile with a side maximum at $\sim 1.3\ a_{0}$. Similarly to what is seen in Fig.~\ref{compositefigure}(b), the LDA $\gamma (\mathbf{r})$'s are well known to overestimate positron annihilation rates in solids. For the 
$\omega=0.4$ system in Fig.~\ref{compositefigure}(b) the two-component LDA of Ref.~\onlinecite{Puska1995} predicts 132~ps whereas the ED result is 261~ps. The zero-positron-density limit of the LDA enhancement gives 115~ps and the GGA 114~ps. 
The discrepancy in the enhancement factors and the lifetime increases with decreasing $\omega$, while the two-component LDA $\sqrt{\gamma (\mathbf{r})}$ of Ref.~\onlinecite{Puska1995} still always reproduces the correct shape.
The LDA assumes a metallic system whereas our system is a finite closed-shell one. On the other hand, the low-density limit of the LDA involves formation of positronium atoms and negative ions, which is not expected in our model system where both particle types are strongly confined by the same potential.


We still need to discuss the results of the various expansions we have presented in terms of how they compare with results of existing models applied in DFT calculations. Figure~\ref{compositefigure}(c) shows normalized spectra divided by the normalized accurate ED spectrum, and compares the convergence of 2-shell diagonal sums of Eq.~(\ref{helikopteri}) corresponding to the sum over pairs of orbitals occupied in the IPM in various bases to that of the natural geminal expansion Eq.~(\ref{geminalmd}), as well as against DFT-based model results. These normalized spectra are more appropriate in theory-experiment comparisons. 
Here we consider the $\omega=0.4$ system but emphasize again that the conclusions drawn do not depend on $\omega$. The normalized 2-shell sums (NO's, LDA and exact KS orbitals) have a common tendency to underestimate the low-momentum intensity and overestimate the high-momentum one. 
On the other hand, Fig.~\ref{compositefigure}(c) further demonstrates the compactness of the natural geminal expansion. The shape of the spectrum is at low momenta very good already with 4 terms and the high-momentum part improves systematically with increasing number of terms.


Of the models applied within DFT, the model of \citet{AlataloPRB1996} uses state-dependent enhancement factors and orbitals calculated with DFT, and the expression used looks similar to the diagonal of Eq.~(\ref{helikopteri}) with the summation restricted to the IPM's states.
When the LDA is used to evaluate the enhancement factors of this model, the results are known to overestimate annihilation with core electrons and thereby the high-momentum intensity.\cite{Makkonen2006} 
This same tendency can be seen in the present results in Fig.~\ref{compositefigure}(c). 
On the other hand, a LDA model with a position-dependent enhancement,\cite{Daniuk1987,*Jarlborg1987} applied using LDA orbitals and the (state-independent) two-component $\gamma (\mathbf{r})$ (Ref.~\onlinecite{Puska1995}), works rather well for our systems concerning especially the shape of the spectrum at low momenta and the high-momentum intensity. The high-momentum part
is oscillatory relative to the reference result unlike other model results. 
These findings are consistent with theory-experiment comparisons, which show that the state-dependent enhancement model overestimates the high-momentum intensity in comparison with experiments and the position-dependent LDA model.\cite{Makkonen2006} However, when a coincindence Doppler spectrum is plotted as a ratio to a reference one, the shape is reproduced better than with the position-dependent LDA model where the oscillations lead to a worse agreement with experiment.
The agreement in the spectra [Fig.~\ref{compositefigure}(c)] and in the extracted enhancement factors [Fig.~\ref{compositefigure}(b)] provide support for the position-dependent enhancement model using an accurate state-independent $\gamma (\mathbf{r})$, although some improvement possibly in the form of state-dependence is needed in order to improve the model. Also one would have to find a way to approximate the $g_{j}$ occupations and better understand the enhancement factors of the low $g_{j}$ geminals, which we leave outside the scope of the present work.

In conclusion, we have introduced an expression for the momentum density of annihilating electron-positron pairs written in terms of
natural geminals that provides both an exact and unique definition for an orthogonal
electron-positron pair wave function familiar from positron literature, and a rapidly convergent
diagonal expression even for strongly interacting
systems. 
The geminals and the concept of enhancement factors can be used as a means to link the problem to the single-particle picture more appropriate for practical modeling more effectively than using a direct orbital expansion of the two-body reduced density matrix.
The natural geminals can be extracted from accurate many-body wave function calculations and used to define and parametrize state and position dependent enhancement factors for models going beyond the independent-particle model and being able to describe correlated electron-positron systems. 
\bibliography{NObibliography}

\end{document}